\documentclass[useAMS,usenatbib]{mn2e}
\usepackage{aas_macros}
\usepackage{graphicx}
\usepackage{hyperref}
\usepackage{enumerate}

\newcommand{\uvot}{\textit{Swift}/UVOT}
\newcommand{\nus}{\textit{NuSTAR}}
\newcommand{\xrt}{\textit{Swift}/XRT}
\newcommand{\bat}{\textit{Swift}/BAT}
\newcommand{\lat}{\textit{Fermi}/LAT}
\newcommand{\ch}{$\chi_{red}^2$}

\newcommand{\po}{power-law}

\newcommand{\lp}{log-parabola}

\newcommand{\bk}{broken power-law}

\newcommand{\nh}{$N_\mathrm{H}$}
\newcommand{\obj}{S5\,0716+714}

\title[First hard X-ray observations of the blazar S5\,0716+714 with NuSTAR]{First hard X-ray observations of the blazar S5\,0716+714 with \textit{NuSTAR} during a multiwavelength campaign}

\author[ A. Wierzcholska and H. Siejkowski ]{
     A. Wierzcholska$^{1}$\thanks{E-mail: alicja.wierzcholska@ifj.edu.pl} and H. Siejkowski$^{2}$\thanks{E-mail: h.siejkowski@cyfronet.pl},\\
     $^{1}$Insitute of Nuclear Physics, Polish Academy of Sciences, ul. Radzikowskiego 152, PL-31-342 Krak\'{o}w, Poland\\
    $^{2}$AGH University of Science and Technology, ACC Cyfronet AGH, ul. Nawojki 11, PO Box 386, PL-30-950, Krak\'{o}w 23, Poland\\
  }

\begin{document}

\date{Accepted .... Received ...; in original form ...}

\pagerange{\pageref{firstpage}--\pageref{lastpage}} \pubyear{2015}

\maketitle

\label{firstpage}

\begin{abstract}
We report the results of a multifrequency campaign targeting \obj\ in the
flaring state of the source observed in 2015 January and February. The observations have been
performed using the following instruments: \textit{Fermi}/Large Area Telescope (LAT), \textit{Nuclear Spectroscopic Telescope Array}, X-ray Telescope and Ultraciolet/Optical Telescope. The
elevated flux level was visible in all frequencies and the outburst consists
of five sub-flares. In this paper  we focus on the analysis of the X-ray observations
both in the soft and hard regimes for data collected with \nus\ and \xrt.  This
is the first time, when hard X-ray observations of the source collected with \nus\ are reported.
The studies reveal both low- and high-energy components
clearly visible in the energy band, with the break energy of 8\,keV,
which is the highest break energy ever reported for \obj.
The second part of this work is concentrated on multifrequency observations collected during
the flaring activity period.  The variability patterns recorded during the period
are characterized using a fractional variability amplitude and
description of the flare profiles. The correlation studies reveal strong and
significant relation between the optical, ultraviolet and $\gamma$-ray
observations, and no time lag is found for any of the studied relations.
\end{abstract}

\begin{keywords}
radiation mechanisms: non-thermal -- galaxies: active -- BL Lacertae objects: general, 
\end{keywords}

\section{Introduction}
Blazars, BL Lacertae (BL Lac) type objects and flat spectrum radio quasars (FSRQs),  are an extreme class of active galactic nuclei, characterized with  polarized and highly variable non-thermal emission observed from the jets pointing at small angles to the observer \citep[e.g.,][]{begelman84}.
The emission is observed in a wide energy range from radio frequencies up to high and very high energy $\gamma$-ray regime \citep[e.g.][]{Vercellone, Abramowski13_1510,   Abramowski2014}.
The spectral energy distribution (SED) of blazars, in $\nu$-$\nu F_{\nu}$ representation, is characterized with a double-bumped shape.
The first, low-energy bump usually is attributed to synchrotron radiation of relativistic electrons from the jet, while the origin of the high-energy bump is still debatable and can be explained in terms of both the leptonic and hadronic scenarios \citep[see e.g.][]{Maraschi92, Mannheim93, Sikora94, Kirk98, Mucke13, Bottcher13}. 
In the most popular so-called Synchrotron-self-Compton model (SSC) this high-energy bump is a result of   the
inverse Compton scattering of the low energy photons collided with the highly energetic leptons.

The position of the low energy peak in blazars SED allows us to distinguish high-, intermediate- and low-energy peaked objects: HBL, IBL, LBL, respectively \citep[see, e.g.,][]{padovani95,fossati98,Abdo2010}. 
For HBL type blazars the low energy peak is situated in the X-ray domain ($\nu_{s}>10^{15}$\,Hz), for IBL blazars in the optical-ultraviolet (UV) range ($10^{14}$\,Hz$<\nu_{s}\leq10^{15}$\,Hz), while in the case of LBL type blazars in the infrared regime  ($\nu_{s}\leq10^{14}$\,Hz) \citep{Abdo2010}.
Different subclasses of blazars FSRQs--LBLs--IBLs--HBLs form a blazar sequence which connects
decreasing bolometric luminosities and $\gamma$-ray dominance in different group of sources \citep{fossati98}.

The blazar \obj\ ($z=0.31$) \citep{Nilsson08}, classified as an IBL type object \citep[e.g][]{Giommi99, Abdo2010}, is one of the brightest and most active blazar. The source was discovered in the 5 GHz Bonn-National Radio Aastronomy Observatory radio survey and included in S5 catalogue \citep{Kuehr81}.
The object was  a target of several optical campaigns focusing on intra-night variability
\citep[e.g.][]{Wagner96, Montagni06, Gupta09, Rani11, Gupta12, Bhatta13, Bhatta15}.

A few instruments monitored the object in the X-ray regime and revealed the spectral upturn, disentangling two spectral  components located in this range \citep[e.g.][]{Ferrero06, Foschini2006, Wierzcholska15}.
The source is included in the first, second and third \lat\ catalogues \citep[1FGL, 2FGL, 3FGL, respectively][]{1FGL, 2FGL, 3FGL} as well as in the \lat\ Bright Source List \citep[0FGL,][]{0FGL}, and the First \lat\ Catalog of Sources Above 10\,GeV \citep[1FHL;][]{1FHL}.
In the very high energy $\gamma$-ray regime \obj\ has been discovered with Major Atmospheric Gamma Imaging Cherenkov Telescopes (MAGIC) telescopes \citep{Anderhub09}.

\cite{Rani2011var} studied multifrequency properties of \obj\ during its  various phases of  activity using optical and radio observations. The authors find higher Doppler factors and higher synchrotron peak frequency for the source, when it is brighter.

Series of radio, optical, X-ray, and  $\gamma$-ray observations of \obj\ collected between 2007 April and 2011 January have been studied by \cite{Rani2013}. The intensive monitoring reveals significant optical variability on time-scales of about 60-70\,d. The optical activity is correlated with the one observed in the $\gamma$-ray regime, which supports SSC mechanism as responsible for the production of the high-energy emission. The optical and $\gamma$-ray emission is also correlated with the radio one, however in the X-ray regime an orphan flare has been observed, which makes one-zone model too simple to explain the activity observed.

Multi-wavelength observations including radio, optical, X-ray and $\gamma$-ray regimes in the studies by \cite{Liao14} show significant variability in all bands. The highest variability amplitudes is observed in the optical and $\gamma$-ray regimes and lower in the X-ray and radio ones. The authors favour the SSC model with the external seed photons originating from the hot dust or broad line region as the best explanation for the emission observed.

In the more recent work, \cite{Chandra15} studied multiwavelength properties of \obj\ during the outburst in 2015 January. The authors report simultaneous optical, X-ray and $\gamma$-ray observations including a time-dependent modelling of the light curves. Furthermore, they find simultaneous variations in all observed bands. The studies support the leptonic origin of the high energy emission observed during the outburst.

This paper focuses on a flaring activity of the source observed in 2015 January-February. 
During the period mentioned the flaring state of the source was reported in different wavelengths: optical range \citep{ATel6957, ATel6944, ATel7004}, near-infrared regime \citep{ATel6902}, X-ray range \citep{Wierzcholska15}  and very high energy $\gamma$ rays regime \citep{ATel6999}.

The paper is organized as follows: Section~\ref{observations} describes the observations and the analysis procedures, Section~\ref{x-ray} is focused on the monitoring of \obj\ in soft and hard X-ray band, in Section~\ref{variability} the multifrequency variability patterns are studied. The work is summarized in Section~\ref{summary}. 

\section{Observations and data analysis}\label{observations}

\subsection{\textit{Fermi}-LAT observations}
The \lat\ is a pair-production satellite telescope sensitive to measure high energy $\gamma$ rays from tens of MeV up to about 500\,GeV \citep{Atwood09}.
The data collected between MJD57023 and MJD57078 in the energy range of 100\,MeV and 300\,GeV were analysed using standard \verb|ScienceToolsv10r0p5| with \verb|P8R2_SOURCE_V6| instrument response function.
For the analysis events within 10$^{\circ}$ region of interest (ROI) centred on \obj\ were selected. The binned maximum-likelihood method \citep{Mattox96} was applied in the analysis.
The Galactic diffuse background was modelled using the \verb|gll_iem_v06| map
cube, and the extragalactic diffuse and residual instrument backgrounds were
modelled jointly using the corresponding isotropic emission template. All sources from the {\it Fermi}-LAT Third Source Catalogue 
\citep{3FGL} inside the ROI of \obj\ were modelled. 

For the spectral analysis  the same energy range as defined earlier was used. To find which model best describes the spectrum, three models: power-law (\verb|PowerLaw|), log-parabola (\verb|LogParabola|) and \verb|PLSuperExpCutoff| has been fitted.  The log-likelihoods of the fits are: $-77938.7$, $-77933.8$ and $-77937.9$, respectively. The comparison of the fit quality to the power-law model, according to the Wilks theorem \citep{Wilks1938}, favours the log-parabola model with test-statistics (TS) equal to 9.7 and \textit{p}-value to 0.002. In the case of the \verb|PLSuperExpCutoff| model the TS is 1.5 and $p\textrm{-value} = 0.214$. Therefore for the further analysis  the log-parabola model is chosen.

\subsection{\textit{NuSTAR} observations}
\begin{figure}
\centering{\includegraphics[width=0.48\textwidth]{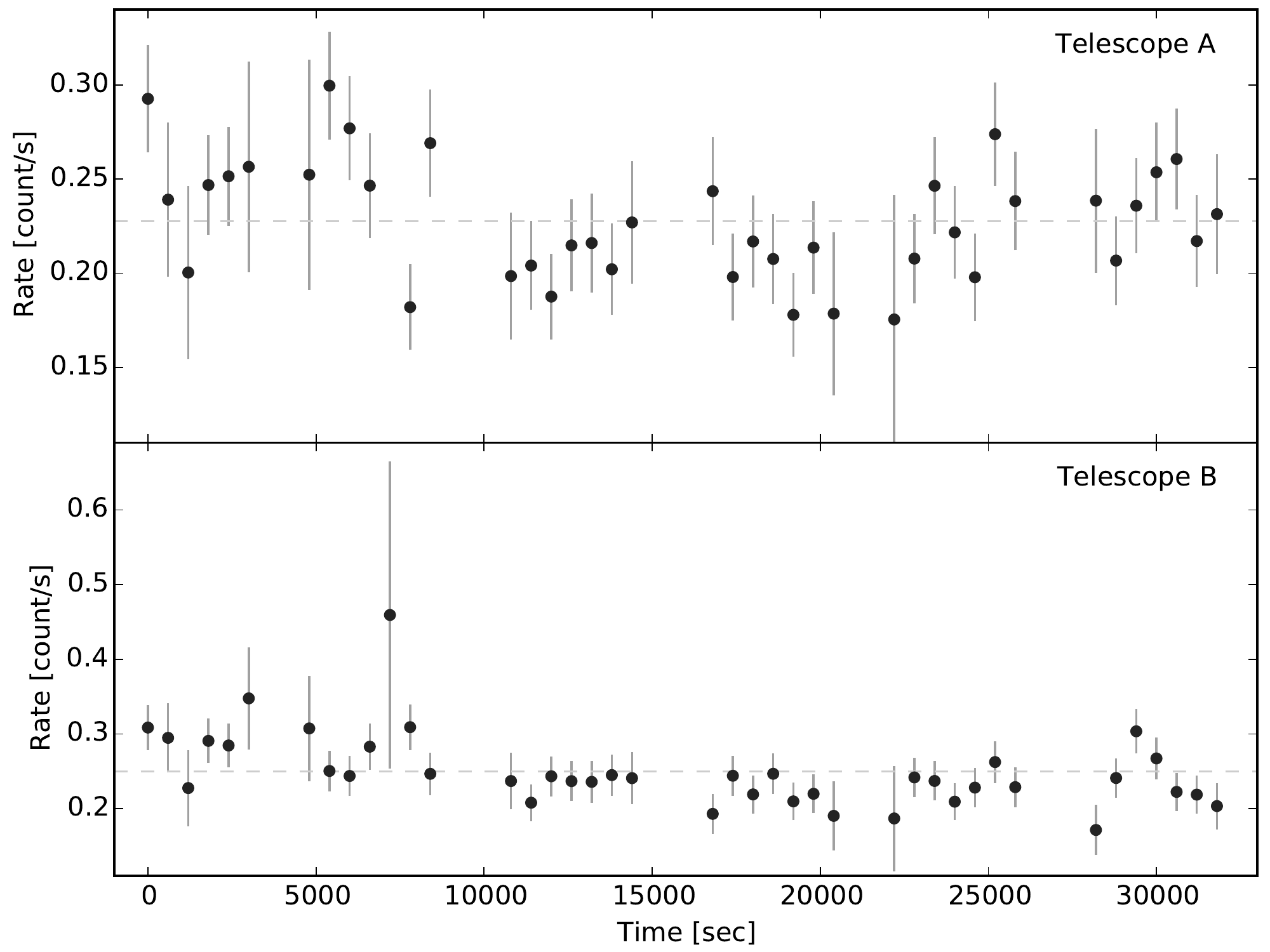}}
\caption{The X-ray count rate light curve measured with \nus. Data points are binned in 600\,s intervals. The upper panel presents data collected with telescope~A, while the bottom one with telescope~B.}
\label{lc_nu}
\end{figure}

The Nuclear Spectroscopic Telescope Array (\textit{NuSTAR}) is a satellite instrument dedicated for observations in the hard X-ray regime (3--79\,keV) \citep{Harrison13}. It consists of two detectors (A and B), with a field of view of each telescope of about 13 arcmin.
\textit{NuSTAR} observed \obj\ twice on 2015 January 24 with the exposures of 338 and 18583\,s (ObsIDs: 90002003001 and 90002003002, respectively).
In both cases the observations were performed in the \verb|SCIENCE| mode.
Due to very short exposure of the first pointing, in the further studies we focus on the second observation only.
The raw data were processed with \nus\ Data Analysis Software package (\textsc{nustardas}) version~1.4.1 (released as a part of \verb|HEASOFT|~6.16) using standard  \verb|nupipeline| task. The data were processed with the most recent calibration version compatible with \textsc{nustardas} v.1.4.1 (version 20140414). 
A source region was selected within a circle with a radius of 0.5 arcmin centred on \obj. The same-size region was selected for a background area. 
In the spectral analysis, we focus on channels which correspond to energy range of 3--40\,keV.  The instrumental response matrices and effective area files were produced with \verb|nuproducts| procedure. 
The count rate light curves for both telescopes  are presented in  Fig.~\ref{lc_nu}.
The subtracted background count rate is $0.228\pm0.005$~count/s and $0.250\pm0.008$~count/s for A and B telescope, respectively.  
Hence, in the spectral fits small normalization factor for the module A with respect to the module B was taken into account. 
Fig.~\ref{lc_nu} does not reveal any significant variability during the observations and the count rate is constant within the error bars.

\subsection{\textit{Swift}-XRT observations}
The Swift Gamma-Ray Burst Mission \citep[hereafter \textit{Swift};][]{Gehrels04} is a multiwavelength space observatory. It is equipped with three instruments: the Burst Alert Telescope \citep[BAT;][]{Barthelmy05}, the X-ray Telescope \citep[XRT;][]{Burrows05}, and the Ultraviolet/Optical Telescope \citep[UVOT;][]{Roming05}.  The \bat\ operates  in the energy range of 15--150\,keV, while \xrt\ in the 0.3--10\,keV range.
\uvot\ observations are provided in siz wavelengths covering the range of 170--600\,nm.

X-ray data in the energy range of 0.3--10\,keV collected with \xrt\ were analysed using version 6.16 of the \textsc{heasoft} package\footnote{\url{http://heasarc.gsfc.nasa.gov/docs/software/lheasoft}}.
Data were recalibrated using the standard procedure \verb|xrtpipeline|.
For the spectral fitting \textsc{xspec} v.12.8.2 was used \citep{Arnaud96}.
All data were binned to have at least 30 counts per bin. 
The light-curve points has been derived by fitting each single observation with a \lp\ model
with a Galactic absorption 
value of $N_{\mathrm{H}} = 3.22 \times 10^{20}$\,cm$^{-2}$ \citep{Kalberla05} set as a frozen parameter.
 The \xrt\ data has been corrected for pile-up effect where needed.

\subsection{\textit{Swift}/UVOT observations}
The UVOT instrument onboard \textit{Swift}  measures the UV and optical emission simultaneously with the X-ray telescope.
The observations are taken in the UV and optical  bands with 
the central wavelengths of: 188\,nm (\textit{UVW}2), 217\,nm (\textit{UVM}2), 251\,nm (\textit{UVW}1), 345\,nm (\textit{U}),
439\,nm (\textit{B}), and 544\,nm (\textit{V}).
The instrumental magnitudes were calculated using \verb|uvotsource| including all photons  from a circular region with a radius of 5 arcsec.
The background was determined from a circular region with a radius of 5 arcsec near the source region, not contaminated with any signal from the nearby sources. 
The flux conversion factors from \cite{Poole08} were used. 
All data were corrected for or the dust absorption using the reddening $E(B-V) = 0.02557$\,mag  \citep{Schlafly} and the ratios of the extinction to reddening, $A_{\lambda} / E(B-V)$, for each filter provided by \cite{Giommi06}.

In order to  estimate the influence of the host galaxy of \obj,  the observations made by
\cite{Nilsson08} are used. They find that the host galaxy has $I$-band magnitude of $17.5 \pm 0.5$ and an effective radius of ($2.7 \pm 0.8$) arcsec. The host galaxy of \obj\ is assumed to be an elliptical galaxy and in order to find the magnitudes in other bands the  templates provided by \cite{Fukugita1995} are used. The host galaxy magnitudes visible by \uvot\ aperture of 5 arcsec are $m_V = 19.6$, $m_B = 21.2$, $m_U = 21.7$. The maximal contribution of the host galaxy to the \uvot\ observations are then: 0.7, 0.3, and $<$0.1 per cent in \textit{V}, \textit{B} and \textit{U} bands, respectively, therefore we find the correction for the host galaxy as negligible.
All the measured magnitudes are collected in Table~\ref{table_uvot}.

\section{Observations in the X-ray regime}\label{x-ray}
During the multifrequency campaign \obj\ was monitored in the X-ray regime with two instruments: \xrt\ and \nus. 
\xrt\ observations with ObsIDs of 00035009154--00035009158 were taken with quasi-simultaneous \nus\ data with ObsID of 90002003002.
For the joint spectral fit, we consider \xrt\ observations taken in the energy range of 0.3--10\,keV and \nus\ observations collected in the energy band of 3--50\,keV. 
We note here that during the period of \xrt\ quasi-simultaneous observations the variability was small. Unfortunately, there are not any ideally simultaneous \xrt\ observations with the \nus\ ones.

Three different models: a single \po, a \bk\ and a \lp,  as defined below, are fitted in order to find the best description of the X-ray observations. 
In each case we include the Galactic hydrogen absorption fixed as a frozen value $N_{\mathrm{H}} = 3.22 \times 10^{20}$\,cm$^{-2}$ \citep{Kalberla05}.
We use following spectral models:
\begin{itemize}
 \item a single power-law:
 \begin{equation}
\frac{dN}{dE}=N_p  \left( \frac{E}{E_0}\right)^{-{\Gamma}},
\end{equation}
 with the spectral index $\Gamma$ and the normalization $N_p$,
\item a logarithmic parabola:
 \begin{equation}
\frac{dN}{dE}=N_l  \left( \frac{E}{E_0}\right)^{-({\alpha+\beta \log (E/E_0)})},
\end{equation}
 with the normalization $N_l$, the spectral parameter $\alpha$ and the curvature parameter $\beta$,

\item a broken power-law:
\begin{equation}
\frac{dN}{dE} = N_b \times\left\{\begin{array}{ll} (E/E_b)^{-\Gamma_1} & \mbox{if $E < E_b$,}\\ (E/E_b)^{-\Gamma_2} & \mbox{otherwise,} \end{array}\right. 
\end{equation}
 \end{itemize}
with the normalization $N_b$, the spectral indices $\Gamma_1$ and $\Gamma_2$ and the break energy $E_b$.

\begin{figure*}
\centering{\includegraphics[width=0.99\textwidth]{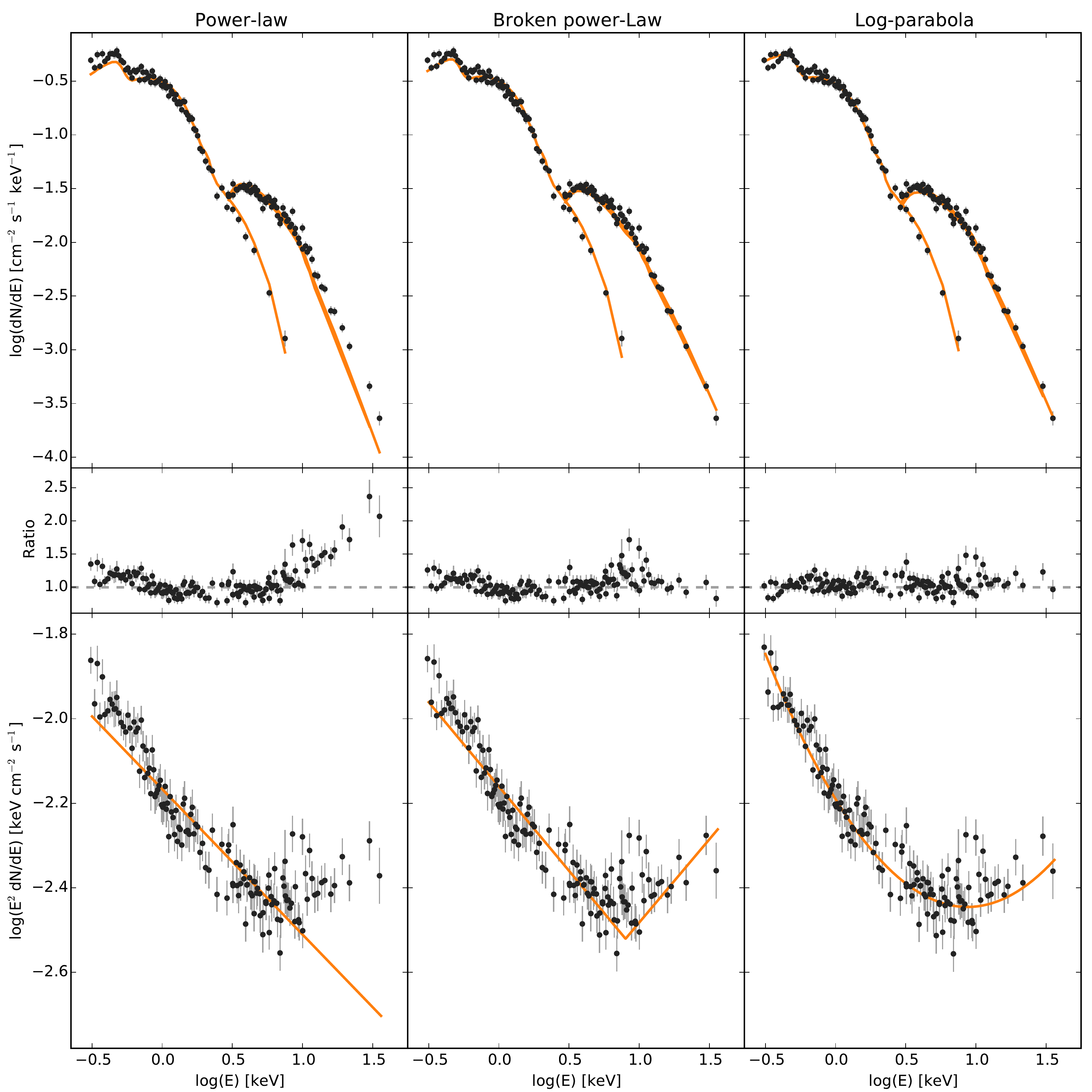}}
\caption{The spectral fits for \obj. The upper panels present the spectral points and the fitted model for the three tested models: \po, \bk,  and \lp.
The middle panel presents the corresponding ratios, while the lower ones show data and the fitted models of the SEDs for \po, \bk, and \lp\ model, respectively. }
\label{spectral_plots}
\end{figure*}

In the case of the \po\ and \lp\ models the scale energy $E_0$ is fixed at 1\,keV. 
Since \xrt\ and \nus\ are not ideally simultaneous, in every case mentioned 2$\%$ of systematic errors are added to the data points.
All the fits parameters are collected in Table~\ref{spectral_fits}, and the spectral fits with residuals and the corresponding SEDs are shown in Fig.~\ref{spectral_plots}.
Based on the \ch\ values and the residual distribution, we conclude that the single \po\ model with a fixed value of the Galactic absorption is not sufficient to describe the X-ray spectrum. We also test the \po\ model with free \nh\ value, which results in lower \ch\ value, i.e. 
$\chi^2_{\mathrm{red}}=1.574$ and $\chi^2_{\mathrm{red}}=1.318$ for the \po\ with Leiden/Argentine/Bonn Survey and free \nh, respectively. The fit with free \nh\ results in  $N_{\mathrm{H}}^\mathit{free} \sim 10^{15}$\,cm$^{-2}$ which is significantly smaller than the one provided by \cite{Kalberla05}. 

The $F$-test \citep[e.g.][]{bevington2003data} shows significant improvement of the \bk\ and \lp\ models relative to the \po\ model, and the corresponding $p$-values are $>10^{-8}$ in both cases. However, it is difficult to distinguish between those two concave models hence we conclude that the \bk\ and \lp\ models both describes the X-ray spectra very well.
It is also worth mentioning here that for both cases: the \bk\ and \lp\ model reveal larger residua around 10\,keV. It is more pronounced in the first case. This may be caused by the fact that the spectrum consists of data collected with two different instruments \xrt\ and \nus. Furthermore, data are not strictly simultaneous and even within \nus\ data set marginal variations are present.

Since the \lp\ is described with a significantly negative curvature, we then conclude that the X-ray regime reveals both the low- and the high-energy spectral components, with the highest break energy value, ever reported for this source, of about of 8\,keV. 

\begin{figure}
\centering{\includegraphics[width=0.48\textwidth]{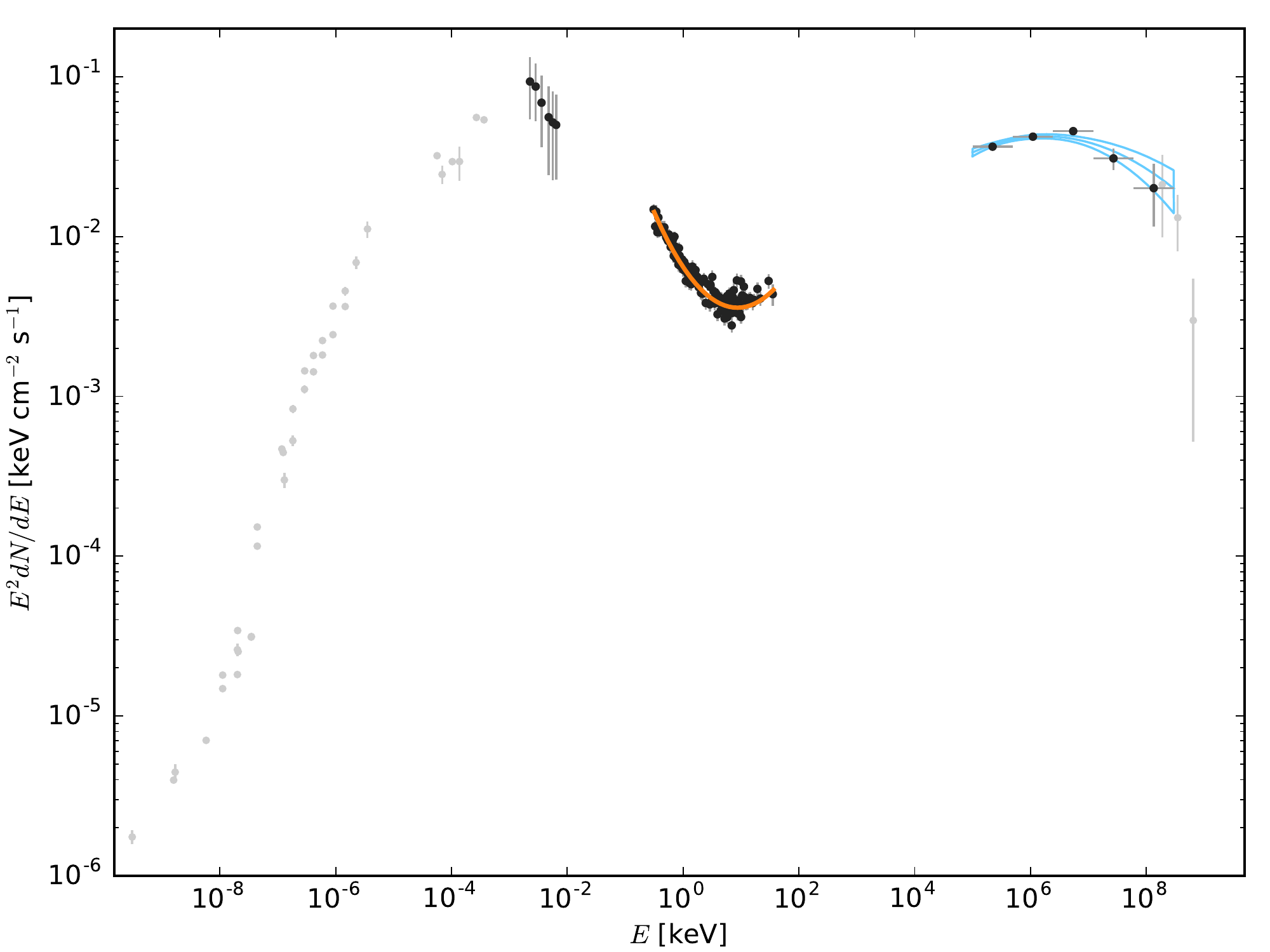}}
\caption{The broad-band SED plot. The black dots show the optical, UV, X-ray and $\gamma$-ray observations collected with \uvot, \xrt, \nus, and \lat. The error bars of the optical data show the standard deviation. The grey points show archive observations (the references are given in text).}
\label{mwl_sed}
\end{figure}

To compare the X-ray spectra with the observation in the optical and $\gamma$-ray regimes  the broad-band SED in the $\nu$-$F_\nu$ representation is shown in Fig.~\ref{mwl_sed}. The optical data are the mean values of the \uvot\ observations and the error bars show the standard deviation. The $\gamma$-ray observations are the \lat\ data and the parameters of the log-parabola fit are following, the normalization is $(2.07 \pm 0.05) \times 10^{-10}$\,cm$^{-2}$\,s$^{-1}$\,MeV$^{-1}$, $\alpha = 1.92 \pm 0.02$ and $\beta = 0.03 \pm 0.01$. The spectral points in the gamma-ray regime are generated by dividing the range analysed into five bins and by fitting the power-law model for each separately. The plot is supplemented with the archive data taken from ASDC\footnote{\url{http://tools.asdc.asi.it/SED/}} and includes observations of the following regimes: radio \citep{Kuehr1981,White1992, Gregory1996, Cohen2007}, infrared \citep{WISE2010,Planck2011,Planck2014} and very high energy (VHE) $\gamma$-ray \citep{Anderhub09}.

\section{Temporal variability}\label{variability}
\begin{figure*}
\centering{\includegraphics[width=0.99\textwidth]{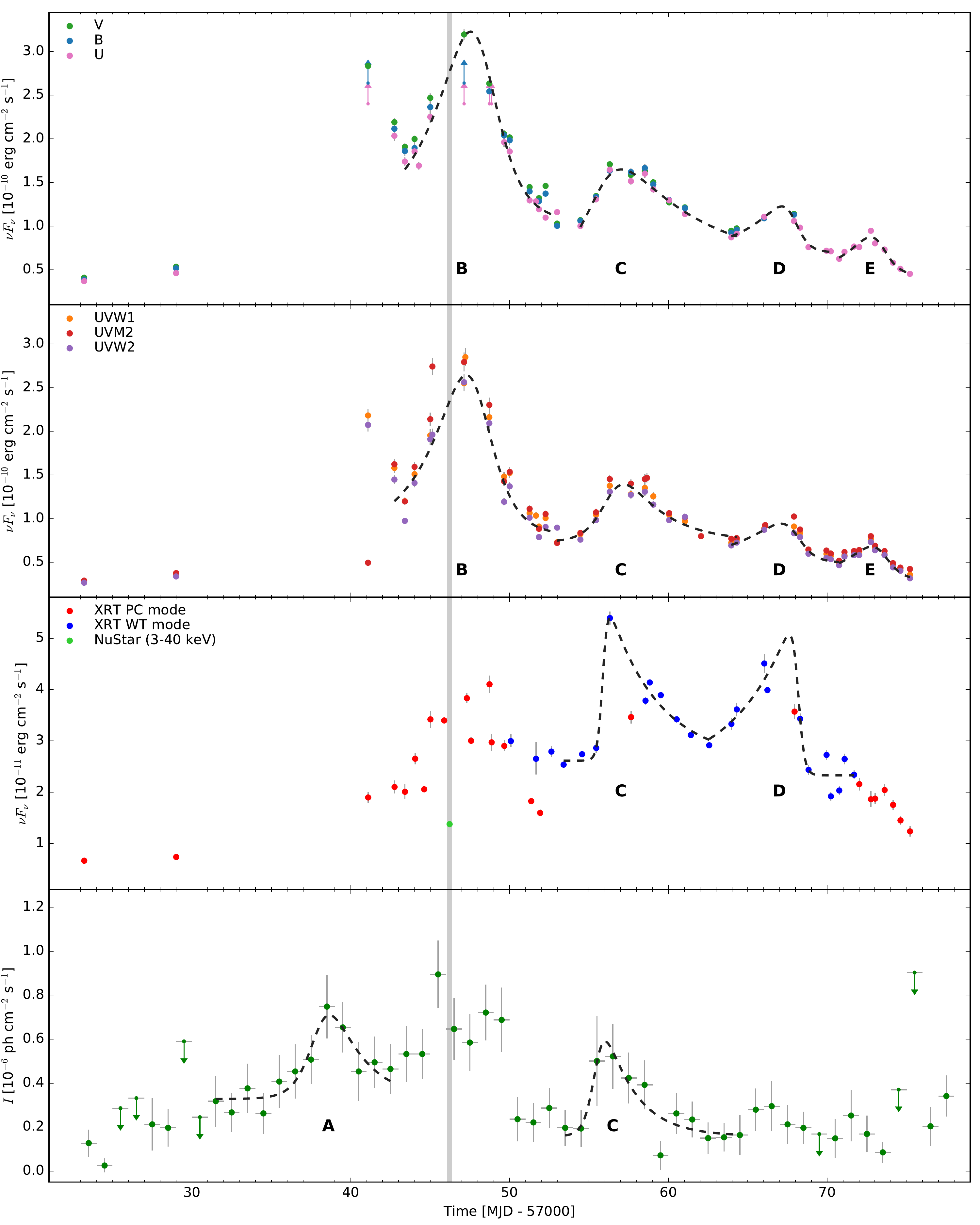}}
\caption{The light curves of \obj. The following panels (from top to bottom) show: the  optical and UV observations with \uvot, the \xrt\ and \nus\ monitoring, and $\gamma$-ray data gathered with \lat. The letters A--E mark the flares defined in Table~\ref{table_flares} and the dashed lines show the fits of the flare profile. The vertical shaded line indicates time of the \nus\ observation.}
\label{lc}
\end{figure*}

The long-term light curve including data collected with \lat, \xrt, \uvot\ is presented in Fig.~\ref{lc}.
The source is active in all the regimes presented.
First look on the light curves suggests that flaring activity of the blazars is different in another regimes. 
These aspects are widely discussed in Section~\ref{correlations}, but it is worth mentioning here that the 
optical observations of \obj\ with \uvot\ in \textit{U}, \textit{B}, and \textit{V} filter are saturated and only lower limits are given in Table~\ref{table_uvot} and presented in Fig.~\ref{lc}. These observations correspond to the highest flux points in the UV bands, which are not saturated. We exclude these optical observations from the further calculations.

\begin{figure}
\centering{\includegraphics[width=0.48\textwidth]{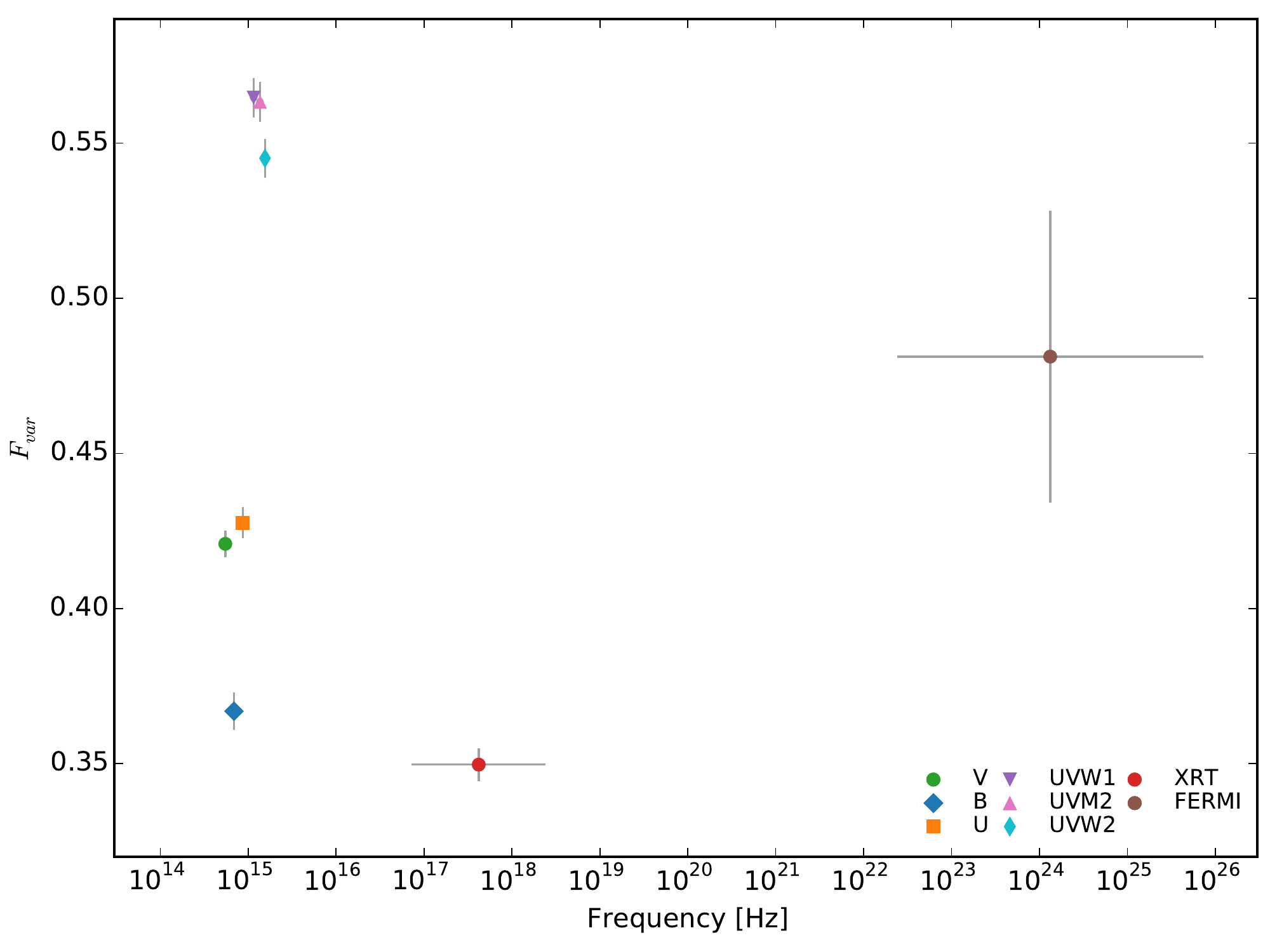}}
\caption{The fractional variability amplitudes for different frequencies.}
\label{ved}
\end{figure}

In order to quantify a temporal variability of \obj, the fractional variability amplitude is calculated following the definition by \cite{Vaughan03}: 
\begin{equation}
 F_\mathrm{var}= \sqrt{\frac{S^2-e^2}{F^2}},
\end{equation}
where $S^2$ is the variance, $e^2$ is the mean square error, and $F$ is the mean flux. 
The uncertainties of $F_\mathrm{var}$ are calculated following the formula by \cite{Poutanen08}:

\begin{equation}
  \delta  F_\mathrm{var}= \sqrt{F_\mathrm{var}^2+(\sigma^2)} -F_\mathrm{var},
\end{equation}
with the error in the normalized excess variance $\sigma$ given as \citep{Vaughan03}
\begin{equation}
 \sigma = \sqrt{\left( \sqrt{\frac{2}{N} }\frac{e^2}{F^2} \right)^2   + \left( \sqrt{\frac{e^2}{N}}\frac{2 F_\mathrm{var}}{F} \right)^2        },
\end{equation}
where $N$ is the number of data points in the light curve. 
Results for different frequencies are collected in Table~\ref{table_fvar} and presented in Fig.~\ref{ved}. The blazar is highly variable in the UV regime. In the optical bands calculated $F_\mathrm{var}$ is lower than for the case of UV observations. We remind here that the highest optical observations are given only as lower limits and excluded from the calculation, hence the lower $F_\mathrm{var}$. 
The lowest $F_\mathrm{var}$ value is found for the X-ray observations.

It must be noted here that $F_\mathrm{var}$ strongly depends on the sampling and size of the time bins. In the case of the current analysis the sampling in \lat\ and \xrt\ are different, i.e. \lat\ has a regular sampling as opposed to \xrt. The size of the time bins influences the $F_\mathrm{var}$ by smoothing out the variability, and lowering its values in the case of larger time bins. Obviously, the time binning is related to the characteristics of the instrument, sensitivity in particular. Another key factor are the flux uncertainties, which according to the definition of $F_\mathrm{var}$ should be constant or at least very close to it. In our analysis this requirement is not strictly fulfilled, however the flux uncertainties are very similar in given energy band.

Since a few significant flares  can be distinguished during time of the outburst, we aim to analyse the individual outbursts in different energy regimes. 
The individual flares are defined in Table~\ref{table_flares}. 
The flux evolution during the flaring event can be characterized using a function, which describes the time profiles of a single flare  \citep{Abdo2010_flares}:
\begin{equation}
\label{flare_profile}
 F(t)=F_e + \frac{F_0}{e^{\frac{t_0-t}{T_\mathrm{r}}}+e^{\frac{t-t_0}{T_\mathrm{d}}}},
\end{equation}
where $F_e$ represents the constant flux level underlying the flare, $F_0$ is the amplitude of the flare,  $t_0$ is the time of the peak and
$T_\mathrm{r}$ and $T_\mathrm{d}$ represent the rise and decay times, respectively.
The time of the peak position $t_\mathrm{m}$ is calculated using the following formula:

\begin{equation}
 t_\mathrm{m}  = t_0 + \frac{T_\mathrm{r} T_\mathrm{d}}{T_\mathrm{r}+T_\mathrm{d}} \ln \left(\frac{T_\mathrm{r}}{T_\mathrm{d}}\right).
\end{equation}
The symmetry of the flare can be described as:
\begin{equation}
 \xi = \frac{T_\mathrm{d} - T_\mathrm{r}}{T_\mathrm{d} + T_\mathrm{r}},
\end{equation}
which can be between $-1$ and 1. The border limits correspond to completely the right and left asymmetric flares, respectively.

In this paper, the fitting procedure is limited to the following rules:  
\begin{enumerate}[(i)]
 \item in the case of the UV observations we focus on four main outbursts visible; fitting procedure is performed on data collected in \textit{UVW}2 band;
 \item in the case of the optical observations we focus on four main outbursts visible; fitting procedure is performed on data collected in \textit{U} band and the upper limit points are not taken into account during the fitting procedure;
 \item in the case of the X-ray observations only the observations taken in Windowed Timing (WT) mode are used, which makes possible to fit only two flares;
 \item in the case of the $\gamma$-ray observations only two significant flares are distinguished and fitted.
\end{enumerate}
The exact dates of the flares selected as well as the fit parameters are listed in Table~\ref{table_flares}.

All of the flares in given wavelengths are asymmetric, and both, right and left, asymmetries are found.
We denote the flares by A--E symbols (also given in Table~\ref{table_flares}) in order to distinguish the simultaneous flares in the further discussion. 
\begin{enumerate}[(i)]
 \item Flare~A is observed only with \lat\ and there is no information about the flux level observed in other wavelengths.
 \item Flare~B is observed with \uvot\ both in the optical and UV bands. The fitted shapes of this outburst in both cases, show weak right asymmetry. 
 \item Flare~C is revealed with \uvot, \xrt, and \lat. In every fit, the shape reveals left asymmetry, and the effect is the strongest for the X-ray data. 
 \item Flare~D is observed in the UV, optical and X-ray regimes. The observations reveal right asymmetry for this outburst.
 \item Flare~E is observed with \uvot\ in the optical and UV regimes. The value of the asymmetry coefficient strongly suggests right asymmetry in the case of this outburst.
\end{enumerate}
The outbursts observed are characterized with not only different asymmetries, but also with different duration times. The duration of the flares, calculated as a sum of the rise and decay times, is between 2.33 and 6.25\,d. It is also worth mentioning that the duration time of a given flare differs between regimes.

\subsection{Correlation studies}
\label{correlations}

To find a relation between the emissions observed on different wavelengths a
cross-correlation function (CCF) is used.  The CCF function is
estimated using the $z$-transformed discrete correlation function (ZDCF)
following the algorithm described in detail by \cite{tal97}. Furthermore a
maximum-likelihood is calculated for each case using the PLIKE algorithm
\citep{tal13} in order to find the peak location for the ZDCF. This peak location,
$\tau_{\mathrm{max}}$, represents the most probable timelag between the two
light curves compared.  For each pair of the light curves the
$\tau_\mathrm{max}$ and the Pearson correlation coefficient for the given light
curves shifted according to the $\tau_\mathrm{max}$ are calculated, as well as
the likelihood of the time lag.
The results are gathered in Table~\ref{table_zdcf}. Every CCF function between the optical and
UV bands give a single and strong likelihood peak for the time lag equals
to zero (within the uncertainties), therefore the values are not reported in
the table.  This can be also verified visually by looking at the light curves in Fig.~\ref{lc}.

\begin{figure}
\centering{\includegraphics[width=0.48\textwidth]{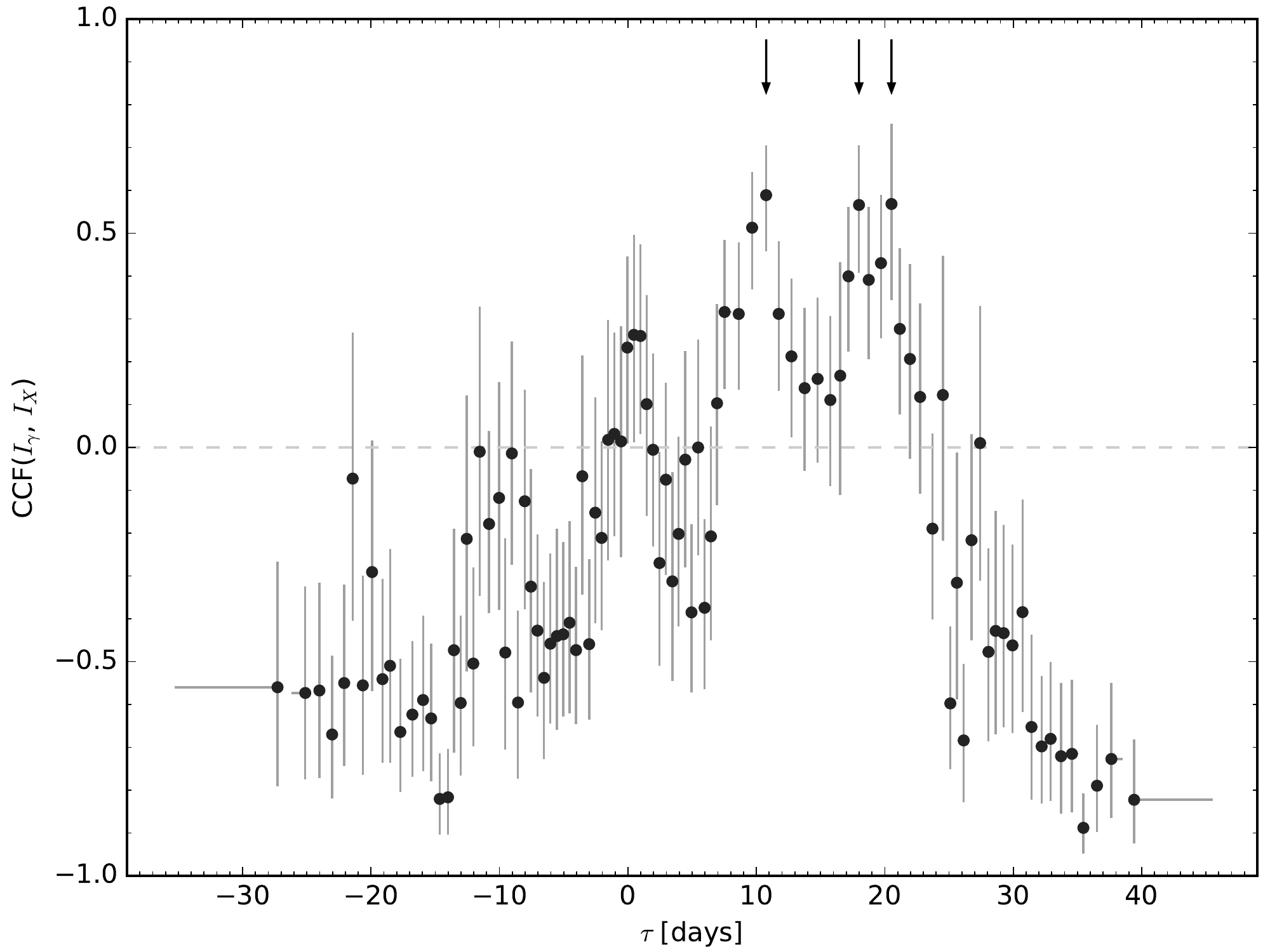}}
\caption{The CCF of the \lat\ ($I_\gamma$) and \xrt\ ($I_X$) light curves
estimated using the ZDCF algorithm. The black arrows marks the maxima reported in Table~\ref{table_zdcf}.}
\label{zdcf_fermi_xrt}
\end{figure}

The cross-correlation of \xrt\ and \lat\ data results in three time lags (see
Fig.~\ref{zdcf_fermi_xrt}). 
The flare B is not well sampled in the X-ray observations. This strongly influences the calculated ZDCF values,
therefore the found time lags of about 11, 18, and 21\,d might be the local
maxima. 
The time distances, between flare B and C is around 10.5\,d and
between B and D is 19.6\,d, and these values correlate with the
time lags calculated.  Assuming that the flare B for the \xrt\ observations
would be more pronounced (similarly as for optical/UV observations), and taking
into account that the maximum flux for \lat\ data is around flare B this could
result in strong correlation for time lag equals to zero or a value close to
it. Therefore the found time lags of 11, 18, and 21\,d could be simply artefacts.

For the comparison of \uvot\ and \lat\ observations, we find two significant
correlations for the \textit{U} and \textit{V} filters and one in the case of the \textit{B} filter.  In
every case, the time lags are with the uncertainties close to
zero days.  Similar situation is in the case of the UV observations. We
do not find any other significant time lag, except for zero days. Please note
that \lat\ data are binned in 1\,d long intervals and therefore the
calculation of the time lags can be slightly disturbed.

\section{Summary and conclusions} \label{summary}
\obj\ is a highly variable blazar and it was monitored in multiple regimes with
different instruments \citep[e.g.][]{Chen08, Liao14, Chandra15}, but in this
paper for the first time, hard X-ray observations collected with \nus are studied.  The paper
focuses on the X-ray monitoring of the blazar during its flaring state in 2015 January.  The
previous soft X-ray campaigns targeting \obj\ has revealed that this regime is a
place, where both  low- and high-energy components meet.  It is also
worth mentioning that during different epochs of observations and different
states of the blazars the break energy shifts between 1.5 and 5.3\,keV
\citep[][]{Tagliaferri03, Donato05, Ferrero06, Wierzcholska15}.

This is the first time, when the soft and hard X-ray observations of \obj\ are studied
together. The joint spectral fit to \xrt\ and \nus\ data confirms a concave shape
and existence of the low- and high-energy components in the
X-ray energy regime.  The shape of the spectrum within the energy range of
0.3--40\,keV is well constrained and the \lp\ or \bk\ models are the
preferable models to describe the X-ray spectrum. The low-energy component is
characterized by a spectral index in the range of about 2.4--2.5, depending on
the model.  We found the break energy -- a point where both components meet --
at about 8\,keV. This is the highest break energy ever reported for \obj.

The second part of this paper is concentrated on the multifrequency observations
of \obj\ with \lat, \nus, \xrt, and \uvot. The flaring activity is observed in
every regime, and five sub-flares can be resolved in different regimes.
The flaring activity of \obj\ is characterized with the fractional variability
amplitudes.  The shape of the flares in different energy regimes is
also characterized with a rise and decay times and asymmetry coefficient.  These
studies have shown that between and within the energy bands the flares are
characterized with different values of the rise and decay times and also
different asymmetry coefficients. We did not find any pattern behaviour for the
 flares described,  but the same flares observed in different energy regimes
have very similar characteristics, including the asymmetry coefficients.  No
significant  shifts between the peak positions of the flares observed in different
wavelengths are found. 

The correlation studies reveal strong positive correlation for
the optical and UV observations for all of the energy band
combinations. We also found such a relation for a comparison of $\gamma$-ray and
optical data, and for $\gamma$-ray and UV data.  In each
case, the calculated value of a time lag is zero days or very close to it.  We
did not find any significant relation for a comparison of the X-ray observations
with other wavelengths, except for the \lat\ data, but the results are probably
artefacts due to the mismatched sampling of the \xrt\ and \lat\ data.

The correlations suggest that the UV, optical and $\gamma$-ray emission observed may
origin from the same emission region.  We have shown that the asymmetry of any
particular flare is the same for different frequencies and the duration of the
flare is shortening with the increasing energy. The results strengthen the hypothesis
that the emission origins from the same region.

%----------------------------------------------------------------

\begin{table*}  
\caption[]{Magnitudes for different epochs from \textit{Swift}/UVOT data for \textit{V}, \textit{B}, \textit{U}, \textit{UVW}1, \textit{UVM}2, and \textit{UVW}2 filters.}
\centering
\begin{tabular}{c|c|c|c|c|c |c|c}
\hline
 Observation ID & Date & \textit{U} & \textit{B} &  \textit{V} &  \textit{UVW}1 & \textit{UVM}2 & \textit{UVW}2     \\
 
 \hline
00035009145 &  2015-01-01 05:04:59  &13.94$\pm$0.05&14.76$\pm$0.05&14.31$\pm$0.05&14.09$\pm$0.06&14.14$\pm$0.06&14.30$\pm$0.06\\
00035009146 &  2015-01-07 00:04:59  &13.70$\pm$0.05&14.45$\pm$0.05&14.02$\pm$0.05&13.84$\pm$0.06&13.86$\pm$0.06&14.04$\pm$0.06\\
00035009147 &  2015-01-19 02:05:48  & $<$11.91     &$<$12.68 &12.21$\pm$0.04&11.87$\pm$0.05&13.56$\pm$0.05&12.07$\pm$0.05\\
00035009148 &  2015-01-20 17:54:59  &12.09$\pm$0.05&12.92$\pm$0.05&12.49$\pm$0.04&12.22$\pm$0.05&12.27$\pm$0.05&12.46$\pm$0.05\\
00035009149 &  2015-01-21 09:44:59  &12.29$\pm$0.05&13.06$\pm$0.05&12.64$\pm$0.04&12.52$\pm$0.05&12.60$\pm$0.05&12.89$\pm$0.05\\
00035009152 &  2015-01-22 00:22:59  &12.19$\pm$0.05&13.04$\pm$0.04&12.59$\pm$0.04&12.27$\pm$0.05&12.29$\pm$0.05&12.49$\pm$0.05\\
00035009153 &  2015-01-22 06:51:59  &12.29$\pm$0.04&--            &--            &--            &--            &--            \\
00035009154 &  2015-01-23 00:08:59  &11.98$\pm$0.05&12.80$\pm$0.05&12.36$\pm$0.04&11.99$\pm$0.05&11.97$\pm$0.05&12.16$\pm$0.05\\
00035009156 &  2015-01-23 03:20:58  &--            &--            &--            &--            &11.70$\pm$0.05&--            \\
00035009157 &  2015-01-25 03:13:59  &$<$11.91      &$<$12.68      &12.08$\pm$0.04&11.70$\pm$0.05&11.68$\pm$0.06&11.84$\pm$0.05\\
00035009158 &  2015-01-25 05:09:59  &--            &--            &--            &11.58$\pm$0.05&--            &--            \\
00035009159 &  2015-01-26 17:21:48  &$<$11.91      &12.72$\pm$0.05&12.29$\pm$0.04&11.88$\pm$0.05&11.89$\pm$0.05&12.06$\pm$0.05\\
00035009160 &  2015-01-26 20:33:58  &$<$11.91      &--            &--            &--            &--            &--            \\
00035009161 &  2015-01-27 15:42:46  &12.13$\pm$0.05&12.96$\pm$0.05&12.56$\pm$0.04&12.29$\pm$0.05&12.41$\pm$0.05&12.67$\pm$0.05\\
00035009162 &  2015-01-28 00:08:58  &12.19$\pm$0.05&12.99$\pm$0.05&12.58$\pm$0.04&12.26$\pm$0.06&12.33$\pm$0.06&12.52$\pm$0.05\\
00035009167 &  2015-01-29 06:13:59  &12.58$\pm$0.05&13.37$\pm$0.05&12.94$\pm$0.04&12.65$\pm$0.05&12.68$\pm$0.05&12.85$\pm$0.05\\
00035009164 &  2015-01-29 15:52:21  &12.59$\pm$0.05&--            &--            &--            &--            &--            \\
00035009168 &  2015-01-29 20:24:58  &12.67$\pm$0.05&13.46$\pm$0.05&13.04$\pm$0.04&12.82$\pm$0.05&12.93$\pm$0.05&13.12$\pm$0.05\\
00035009169 &  2015-01-30 06:21:59  &12.76$\pm$0.04&13.39$\pm$0.04&12.93$\pm$0.04&12.71$\pm$0.05&12.74$\pm$0.05&12.97$\pm$0.05\\
00035009170 &  2015-01-30 23:47:59  &12.70$\pm$0.04&13.73$\pm$0.04&13.31$\pm$0.04&13.06$\pm$0.05&13.15$\pm$0.06&12.98$\pm$0.05\\
00035009171 &  2015-02-01 10:59:58  &12.86$\pm$0.05&13.67$\pm$0.04&13.27$\pm$0.04&12.93$\pm$0.05&12.99$\pm$0.05&13.16$\pm$0.05\\
00035009172 &  2015-02-02 10:38:59  &12.57$\pm$0.05&13.42$\pm$0.05&13.02$\pm$0.04&12.67$\pm$0.05&12.72$\pm$0.05&12.88$\pm$0.05 \\
00035009173 &  2015-02-03 07:23:59  &12.32$\pm$0.05&13.20$\pm$0.05&12.76$\pm$0.04&12.37$\pm$0.05&12.39$\pm$0.06&12.57$\pm$0.05\\
00035009174 &  2015-02-04 15:17:20  &12.41$\pm$0.05&13.21$\pm$0.05&12.84$\pm$0.04&12.45$\pm$0.05&12.43$\pm$0.05&12.60$\pm$0.05\\
00035009175 &  2015-02-05 12:23:59  &12.35$\pm$0.05&13.18$\pm$0.05&12.81$\pm$0.04&12.39$\pm$0.05&12.39$\pm$0.06&12.57$\pm$0.05\\
00035009176 &  2015-02-05 15:34:59  &--            &--            &--            &--            &12.38$\pm$0.05&--            \\
00035009177 &  2015-02-06 01:09:59  &12.48$\pm$0.05&13.31$\pm$0.04&12.90$\pm$0.04&12.47$\pm$0.05&--            &12.70$\pm$0.05\\
00035009178 &  2015-02-07 01:06:00  &12.58$\pm$0.04&13.45$\pm$0.04&13.08$\pm$0.04&12.67$\pm$0.05&12.73$\pm$0.05&12.88$\pm$0.05\\
00035009179 &  2015-02-08 00:55:00  &12.72$\pm$0.05&13.53$\pm$0.04&13.13$\pm$0.04&12.75$\pm$0.05&12.78$\pm$0.05&12.84$\pm$0.05\\
00035009180 &  2015-02-09 01:15:59  &--            &--            &--            &--            &13.04$\pm$0.06&--            \\
00035009181 &  2015-02-10 22:59:59  &13.01$\pm$0.05&13.82$\pm$0.05&13.40$\pm$0.04&13.07$\pm$0.05&13.08$\pm$0.06&13.26$\pm$0.05\\
00035009182 &  2015-02-11 07:09:58  &12.96$\pm$0.05&13.78$\pm$0.05&13.37$\pm$0.04&13.02$\pm$0.06&13.07$\pm$0.06&13.21$\pm$0.05\\
00035009184 &  2015-02-13 00:42:30  &12.75$\pm$0.05&13.64$\pm$0.05&--            &12.85$\pm$0.05&--            &13.01$\pm$0.05\\
00035009185 &  2015-02-13 02:08:43  &--            &--            &-- &-- &12.88$\pm$0.05&-- \\
00035009186 &  2015-02-13 02:24:59  &--            &--            &-- &-- &12.89$\pm$0.05&--  \\
00035009190 &  2015-02-14 21:45:05  &12.80$\pm$0.05&13.60$\pm$0.05& 13.20$\pm$0.04&12.82$\pm$0.05&12.77$\pm$0.05&13.06$\pm$0.07\\
00035009187 &  2015-02-15 06:54:58  &12.88$\pm$0.05&--            &--               &13.91$\pm$0.05&12.94$\pm$0.06&13.12$\pm$0.05\\
00035009188 &  2015-02-15 19:34:59  &13.16$\pm$0.05&--            &--               &13.20$\pm$0.06&13.27$\pm$0.06&13.42$\pm$0.05\\
00035009189 &  2015-02-16 22:45:58  &13.22$\pm$0.05&--            &--               &13.28$\pm$0.05&13.29$\pm$0.06&13.51$\pm$0.05\\
00035009192 &  2015-02-17 05:21:59  &13.23$\pm$0.05&--            &--               &13.32$\pm$0.05&13.35$\pm$0.06&13.54$\pm$0.05\\
00035009193 &  2015-02-17 17:58:59  &13.37$\pm$0.05&--            &--               &13.48$\pm$0.06&13.51$\pm$0.06&13.69$\pm$0.06\\
00035009194 &  2015-02-18 02:02:59  &13.24$\pm$0.05&--            &--               &13.31$\pm$0.06&13.32$\pm$0.06&13.48$\pm$0.05\\
00035009195 &  2015-02-18 16:24:59  &13.15$\pm$0.05&--            &--               &13.27$\pm$0.05&13.30$\pm$0.06&13.45$\pm$0.05\\
00035009196 &  2015-02-19 00:14:59  &13.16$\pm$0.05&--            &--               &13.23$\pm$0.06&13.28$\pm$0.06&13.45$\pm$0.05\\
00035009197 &  2015-02-19 17:54:59  &12.92$\pm$0.05&--            &--               &13.01$\pm$0.06&13.04$\pm$0.06&13.20$\pm$0.06\\
00035009198 &  2015-02-20 00:10:59  &13.10$\pm$0.05&--            &--               &13.16$\pm$0.05&13.20$\pm$0.06&13.35$\pm$0.05\\
00035009199 &  2015-02-20 14:32:59  &13.20$\pm$0.05&--            &--               &13.29$\pm$0.06&13.30$\pm$0.06&13.45$\pm$0.05\\
00035009200 &  2015-02-21 03:18:59  &13.45$\pm$0.05&--            &--               &13.53$\pm$0.06&13.57$\pm$0.06&13.75$\pm$0.05\\
00035009201 &  2015-02-21 14:29:59  &13.59$\pm$0.05&--            &--               &13.65$\pm$0.06&13.69$\pm$0.06&13.85$\pm$0.05\\
00035009202 &  2015-02-22 04:56:58  &13.72$\pm$0.05&--            &--               &13.84$\pm$0.06&13.73$\pm$0.06&14.11$\pm$0.06\\
\hline

\end{tabular}
%\tablefoot{}
\label{table_uvot}
\end{table*}

\begin{table*}  
\caption[]{Results of the joint spectral fits to the \xrt\ and \nus\ observations in the X-ray range.  The following columns present: (1)  the chosen model: \po, \bk\ or \lp; (2) the normalization given in 10$^{-3}$\,cm$^{-2}$\,s$^{-1}$\,keV$^{-1}$;  (3) the photon index for the \po\ and \lp\ model, or the low-energy photon index for the \bk\ model; (4) the high-energy photon index for the \bk\ model, or the curvature parameter for the \lp\ model; (5) the break energy for the \bk\ model given in keV; (6) the unabsorbed model flux in the energy range of 2--10\,keV; (7) the unabsorbed model flux in the energy range of 10--20\,keV; (8) the unabsorbed model flux in the energy range of 20--50\,keV; (9) the reduced $\chi^2$ value and the number of degrees of freedom. The values for (6--8) are given in 10$^{-12}$\,erg\,cm$^{-2}$\,s$^{-1}$.}
\centering
\begin{tabular}{l|c|c|c|c|c|c|c|c}
\hline
Model & Normalization & $\Gamma$/$\Gamma_1$/$\alpha$ & $\Gamma_2$/$\beta$  & E$_{\mathrm{br}}$  & F$_{2-10}$  & F$_{10-20}$  & F$_{20-50}$ & $\chi^2$(dof)    \\ 
(1) & (2) & (3) & (4)  & (5)  & (6)  & (7)  & (8) & (9) \\ 

 \hline
\po\ & $6.815\pm0.068$ & $2.344\pm0.008$ & --               & --            & $10.58\pm0.10$ & $3.04\pm0.02$ & $3.05\pm0.05$ & 1.574(513) \\
\bk\ & $6.912\pm0.067$ & $2.399\pm0.009$ &  $1.606\pm0.053$ & $8.01\pm0.56$ & $10.03\pm0.10$ & $4.19\pm0.02$ & $7.64\pm0.37$ & 1.197(511) \\
\lp\ & $6.450\pm0.067$ & $2.543\pm0.013$ & $-0.289\pm0.016$ & --            & $10.04\pm0.01$ & $4.13\pm0.16$ & $6.58\pm0.29$ & 1.041(512) \\

\hline

\end{tabular}
%\tablefoot{}
\label{spectral_fits}
\end{table*}

\begin{table*}  
\caption[]{The fractional variability for different energy bands. The following columns present: (1) the name of the instrument, (2) the energy band or filter, (3) the fractional variability, and (4) the $\chi^2$ value and the number of degrees of freedom for the fit with a constant.}
\centering
\begin{tabular}{c|c|c|r@{/}l}
\hline
Instrument & Energy band/filter & $F_{\mathrm{var}}$ &\multicolumn{2}{c}{$\chi^2$/$n_{\mathrm{dof}}$} \\
(1) & (2) & (3) & \multicolumn{2}{c}{(4)} \\

\hline
\lat\    & 0.1 -- 500\,GeV & $ 0.4812\pm0.0471 $ &    288 & 48 \\

\xrt\    & 2 -- 10\,keV    & $ 0.3497\pm0.0053 $ & 8\,562 & 60 \\

\uvot    & \textit{UVW}2            & $ 0.5451\pm0.0063 $ & 18\,193 & 42 \\

\uvot    & \textit{UVM}2            & $ 0.5633\pm0.0064 $ & 14\,800 & 44 \\

\uvot    & \textit{UVW}1            & $ 0.5646\pm0.0063 $ & 18\,233 & 43 \\

\uvot    & \textit{U}               & $ 0.4277\pm0.0050 $ & 17\,836 & 40 \\

\uvot    & \textit{B}               & $ 0.3669\pm0.0061 $ & 14\,572 & 25 \\

\uvot    & \textit{V}               & $ 0.4209\pm0.0044 $ & 27\,355 & 26 \\

\hline
\end{tabular}
\label{table_fvar}
\end{table*}

\begin{table*}
\caption[]{The identified flares in the light curves of \textit{U}, \textit{UVW}2, $\gamma$-ray and X-ray bands, and the fit parameters of the flare profile (equation \ref{flare_profile}). The following columns present: (1) the studied energy band; (2) the time interval of a flare; (3) the flare symbol; (4) the approximate time of the peak (5) the rise time; (6) the decay time; (7) the calculated time of a peak; (8) the symmetry coefficient.}
 \begin{tabular}{c|c|c|c|c|c|c|c}
 \hline
 Energy band & Time interval & Flare symbol & $t_0$ & $T_\mathrm{r}$ & $T_\mathrm{d}$ & $t_\mathrm{m}$ & $\xi$ \\
 (1)  & (2)  & (3)  & (4) & (5) & (6) & (7) & (8)\\
             & (d)        &              & (MJD-57000) & (d) & (d) & (MJD-57000) & \\
 \hline
\uvot(\textit{U})  & 43.4-53.0 & B & $48.15 \pm  0.66$ & $ 2.47 \pm  0.82$ & $1.15 \pm 0.58$ & 47.55 & $-0.37$ \\
\uvot(\textit{U})  & 54.5-64.3 & C & $55.64 \pm  0.66$ & $ 1.03 \pm  0.25$ & $5.22 \pm 2.69$ & 57.03 & $0.67$ \\
\uvot(\textit{U})  & 64.0-70.2 & D & $67.82 \pm  0.20$ & $ 2.59 \pm  0.63$ & $0.49 \pm 0.14$ & 67.14 & $-0.68$ \\
\uvot(\textit{U})  & 70.7-75.2 & E & $73.27 \pm  0.39$ & $ 1.94 \pm  0.73$ & $0.58 \pm 0.29$ & 72.73 & $-0.54$ \\
\uvot(\textit{UVW}2) & 42.7-53.0 & B & $47.84 \pm  0.70$ & $ 2.34 \pm  0.64$ & $1.09 \pm 0.46$ & 47.27 & $-0.36$ \\
\uvot(\textit{UVW}2) & 53.0-64.3 & C & $56.46 \pm  0.61$ & $ 0.80 \pm  0.42$ & $2.60 \pm 1.19$ & 57.18 & $0.53$ \\
\uvot(\textit{UVW}2) & 64.0-70.7 & D & $67.93 \pm  0.29$ & $ 3.23 \pm  0.95$ & $0.61 \pm 0.22$ & 67.08 & $-0.68$ \\
\uvot(\textit{UVW}2) & 70.7-75.2 & E & $73.41 \pm  0.41$ & $ 2.21 \pm  0.86$ & $0.60 \pm 0.33$ & 72.80 & $-0.57$ \\
\xrt\     & 53.4-62.6 & C & $55.88 \pm  0.42$ & $ 0.17 \pm  0.14$ & $3.15 \pm 0.97$ & 56.34 & $0.90$ \\
\xrt\     & 62.6-71.7 & D & $68.15 \pm  0.21$ & $ 3.58 \pm  1.28$ & $0.20 \pm 0.25$ & 67.60 & $-0.89$ \\
\lat\     & 31.5-42.5 & A & $38.11 \pm  0.76$ & $ 0.77 \pm  0.48$ & $2.05 \pm 1.00$ & 38.66 & $0.45$ \\
\lat\     & 53.5-64.5 & C & $55.50 \pm  0.35$ & $ 0.39 \pm  0.30$ & $1.94 \pm 1.01$ & 56.02 & $0.67$ \\

\hline
\end{tabular}
\label{table_flares}
\end{table*}

\begin{table*}
\caption[]{The maxima of the CCF distribution. The following columns present: (1) the instruments used in the calculation;  (2) the calculated value of the time lag; (3) the Pearson correlation coefficient for a given time lag, and (4) the likelihood of the given time lag.}
\centering
\begin{tabular}{c|c|c|c}
\hline

Instruments & $t_\mathit{lag}$ (d) & $R$ & Likelihood \\
(1) & (2) & (3) & (4) \\

\hline
\lat--\uvot(\textit{V})    & $1.31_{-0.29}^{+0.23}$  & $0.88_{-0.06}^{+0.05}$  & 0.0331  \\
\lat--\uvot(\textit{V})    & $0.22_{-0.26}^{+0.24}$  & $0.88_{-0.08}^{+0.06}$  & 0.0306  \\
\lat--\uvot(\textit{B})    & $-0.04_{-0.19}^{+0.19}$ & $0.88_{-0.08}^{+0.06}$  & 0.0215  \\
\lat--\uvot(\textit{U})    & $-0.13_{-0.10}^{+0.15}$ & $0.90_{-0.06}^{+0.05}$  & 0.0199  \\
\lat--\uvot(\textit{U})    & $0.92_{-0.13}^{+0.18}$  & $0.91_{-0.06}^{+0.04}$  & 0.0244  \\
\lat--\uvot(\textit{UVW}1) & $1.06_{-0.29}^{+0.19}$  & $0.84_{-0.08}^{+0.07}$  & 0.0184  \\
\lat--\uvot(\textit{UVM}2) & $0.22_{-0.08}^{+0.03}$  & $0.90_{-0.07}^{+0.05}$  & 0.0517  \\
\lat--\uvot(\textit{UVW}2) & $0.47_{-0.26}^{+0.07}$  & $0.85_{-0.07}^{+0.06}$  & 0.0712  \\
\lat--\xrt        & $10.76_{-0.29}^{+0.39}$ & $0.59_{-0.13}^{+0.12}$  & 0.0129  \\
\lat--\xrt        & $17.99_{-0.19}^{+0.20}$ & $0.57_{-0.16}^{+0.14}$  & 0.0114  \\
\lat--\xrt        & $20.52_{-0.06}^{+0.21}$ & $0.57_{-0.22}^{+0.19}$  & 0.0151  \\
\hline

\end{tabular}
\label{table_zdcf}
\end{table*}

\section*{Acknowledgements}
The authors thank anonymous referee for suggestions and remarks provided.
This research was supported in part by PLGrid Infrastructure.
The plots presented in this paper are rendered using \textsc{matplotlib} \citep{matplotlib}.

\bibliographystyle{mn2e_williams}
\bibliography{references}

\label{lastpage}
\end{document}